# Multiphonon Raman Scattering in Graphene

Rahul Rao,[1][*][†] Derek Tishler,[2] Jyoti Katoch,[2] and Masa Ishigami[2]

[1]Materials and Manufacturing Directorate, Air Force Research Laboratory, WPAFB, OH 43433, USA
[2]Department of Physics and Nanoscience Technology Center, University of Central Florida, Orlando, FL 32816, USA

ABSTRACT

We report multiphonon Raman scattering in graphene samples. Higher order combination modes involving 3 phonons and 4 phonons are observed in single-layer (SLG), bi-layer (BLG), and few layer (FLG) graphene samples prepared by mechanical exfoliation. The intensity of the higher order phonon modes (relative to the G peak) is highest in SLG and decreases with increasing layers. In addition, all higher order modes are observed to upshift in frequency almost linearly with increasing graphene layers, betraying the underlying interlayer van der Waals interactions.

[*] Coresponding Author, Email: rrao@honda-ri.com
[†] Present Address: Honda Research Institute, 1381 Kinnear Rd., Columbus, OH.

Single to multilayer graphene display a variety of unusual electronic and transport properties such as the Dirac physics, tunable band gap due to the broken symmetry by additional layers, and electron interaction effects.[1] Light-interaction with graphene is highly important for both graphene science and technology as optical techniques like Raman spectroscopy reveal fundamental properties of graphene such as doping levels, defect concentrations,[2] and the utility of graphene in optoelectronics seems promising. Raman spectroscopy is also the standard technique for characterizing graphene samples due to distinct features that depend strongly on the number of layers,[3,4] as well as the stacking order in few-layer graphene.[5-8] In particular, the second order 2D (also called the G') peak, which occurs due to a double resonance Raman process involving *inter*-valley scattering of an electron by two transverse optical (iTO) phonons, is the most intense feature in the Raman spectrum of single layer graphene (SLG) on $SiO_2$ and can be fitted with a single Lorentzian peak.[2,3,9] On the other hand, the 2D peak in bilayer graphene (BLG) is composed of four Lorentzian peaks, and reflects the hyperbolic electronic band structure due to the stacking between two graphene layers. As the number of layers increase, the 2D peak evolves into a two-peak structure due to coupling between graphene layers in a three dimensional crystal.[2,10] Furthermore, the intensity of the 2D peak diminishes with respect to the G peak with increasing graphene layers, and has been used to identify the number of layers in graphene samples.[11]

While the two-phonon 2D peak has received much attention in graphene samples, higher order modes involving multiple phonons remain unexplored. Multi-phonon Raman scattering is generally weaker in bulk materials due to a vanishing DOS for higher order phonons.[12] Yet, multiphonon Raman scattering can be observed in single- (SWNTs),[13,14] multi-walled carbon nanotubes (MWNTs),[15] and highly oriented pyrolytic graphite (HOPG).[14, 16] Wang *et al.*[13] reported intense higher order combination modes involving up to 6 phonons (between 2500 – 8500 cm$^{-1}$) in individual SWNTs, which was made possible by large resonance enhancements due to coupling with the excitation laser. On the other hand, the multiphonon modes in MWNTs[15] and HOPG[14, 16] are much weaker in intensity compared to their one-phonon or two-phonon modes, and are difficult to observe. In the same vein, it is of considerable interest to determine whether such higher order modes can also be observed in graphene, and how they evolve with increased layer stacking. Here we show higher order combination modes (up to 4 phonons) from single and multiple graphene layers prepared by exfoliation from HOPG on SiO$_2$ substrates.[17] We find that the multiphonon modes are most intense in single layer graphene (SLG) and decrease in intensity with increasing layers. In addition, the three-phonon modes are observed to upshift in frequency with increasing number of layers, presumably due to van der Waals interactions caused by layer stacking.

Raman spectra ($E_{laser}$ = 2.33 eV) from single-layer (SLG), bi-layer (BLG), and few-layer (FLG) graphene samples are shown along with a spectrum from HOPG in Fig. 1. The first order E$_{2g}$ mode (G peak) at ~1585 cm$^{-1}$ and the overtone of the iTO phonon mode (2D peak) at ~2700 cm$^{-1}$ exhibit frequencies and lineshapes similar to what have been described previously;[3, 5] the 2D peaks in SLG, BLG and FLG can be fitted with 1, 4, and 2 Lorentzian peaks, respectively. Beyond the 2D peak frequency, several weak intensity modes can be observed between 3000 – 6000 cm$^{-1}$. The sharp peak (single Lorentzian) at ~3230 cm$^{-1}$ is an overtone of the D' peak and is called the 2D' peak. The D' peak is a disorder-induced peak occurring at ~1620 cm$^{-1}$ in sp$^2$ carbon samples, and is caused by double resonance *intra*-valley scattering of a photo-excited electron by a phonon along with elastic scattering by a defect. The 2D' peak at ~3230 cm$^{-1}$ is its overtone and like the 2D peak, does not need defects for activation. The overtone of the G peak (2G) at ~3160 cm$^{-1}$ is generally not observed in the graphite Raman spectrum,[18] although it has been observed previously in the resonance Raman spectra from SWNTs.[13] Hence as described below, we assign some of the higher order modes above 4500 cm$^{-1}$ to combinations involving the 2D' peak, rather than the 2G peak as assigned by earlier reports.[15, 16]

Beyond 4000 cm$^{-1}$, a peak at ~4250 cm$^{-1}$ is the most intense feature in the Raman spectra shown in Fig. 1. This feature is assigned to a combination of the G and 2D modes (G+2D) as explained further below.[19] The intensity of the G+2D peak is highest for SLG and decreases in intensity with increasing layers, tracking the intensity of the 2D peak, which is also the most intense for SLG compared to multi-layered graphene. A weak intensity peak at ~4030 cm$^{-1}$ is also observed in all samples and is assigned to a combination of the D and 2D modes (D+2D). Both the G+2D and D+2D peaks are shown more clearly in Fig. 2 where the spectra have been normalized by the intensity of the G+2D peak and fitted with Lorentzian peaks. The G+2D peak in SLG can be fit with a single Lorentzian peak; hence its lineshape is similar to that of the 2D peak in SLG. However, the linewidth of the G+2D peak (FWHM ~ 85 cm$^{-1}$) is greater than the 2D peak (FWHM ~30 cm$^{-1}$). On the other hand, the G+2D peaks in BLG, FLG, and HOPG can be deconvoluted into multiple peaks, reflecting the changes in the electronic band structure brought about by coupling between graphene layers; this is also observed in the 2D peak from BLG and FLG samples.[2, 3, 9] The assignments of the G+2D and D+2D peaks can be confirmed by the dispersion of the peaks with increasing laser energy ($E_{Laser}$ = 2.33, 2.54, and 3.81 eV), as shown in Fig. 3a. The dispersion of the G+2D peak is ~100 cm$^{-1}$/eV, which should be similar to the dispersion of the 2D peak (~95 cm$^{-1}$/eV) since the G peak is dispersionless. On the other hand, the dispersion of the D+2D peak should approximately equal the sum of the dispersions of the D (typically ~50 cm$^{-1}$/eV)[20] and 2D peaks (~95 cm$^{-1}$/eV), and is found to be ~130 cm$^{-1}$/eV. Similar dispersions can also be observed for the G+2D peak in BLG, as shown in Fig. 4. The G+2D peak in BLG can be deconvoluted into 4 peaks, similar to the 2D peak in BLG[3, 9]. The dispersions of the four components within the 2D peak have been reported to vary between 80 - 100 cm$^{-1}$/eV.[21-23] As shown in Fig. 4, the dispersion of the highest frequency component within the G+2D peak in BLG is ~98 cm$^{-1}$/eV, and the dispersions of the other peaks within the G+2D as well as 2D peaks are similar. Moreover, the differences in frequencies between the four components of the G+2D peak (~20-40 cm$^{-1}$) are similar to those in between the four components in the 2D peak in BLG.

Another interesting feature that can be observed in the spectra shown in Fig. 2 (indicated by the dotted lines) is that the G+2D and D+2D peaks appear to upshift in frequency with increasing graphene layers. Such upshifts with increasing graphene layers have been observed recently for combination modes between 1700 – 2300 cm$^{-1}$ involving optical and acoustic phonons, as well as for the G peak phonons in exfoliated graphene.[5] As shown in the plot of peak frequencies versus number of layers (1/$n$) in Fig. 3b, the G+2D and D+2D peaks exhibit an almost linear dependence on 1/$n$ and the data can be fitted by an equation of the form $\omega(n) = \beta/n + \omega(\infty)$, where $n$ is the number of graphene layers, and $\beta$ is a constant.[4] The values of $\beta$ (50 cm$^{-1}$ and 25 cm$^{-1}$ for the G+2D and D+2D peaks respectively) obtained from the linear fits in Fig. 3b are comparable to those reported for combination modes in graphene.[5] Frequency upshifts of the G band in SLG (up to 13 cm$^{-1}$) have been found to occur due to unintentional doping.[24] However, in addition to an upshift in the G peak frequency, there is a corresponding narrowing of the G peak and decrease in the 2D peak intensity with unintentional (and inhomogeneous) doping.[24] In all the spectra used in this study we have confirmed the consistency of the G peak linewidth (for example, FWHM ~ 9-11 cm$^{-1}$ in SLG) as well as the ratio of intensities between the 2D and G peaks between all samples and multiple spots measured on each sample. The frequency upshifts shown in Fig 3b are likely due to van der Waals interactions in layered systems and, in fact, have been observed recently in the phonon modes of few-layered $MoS_2$ samples.[25]

Finally, we turn our attention to several weak intensity modes observed above 4500 cm$^{-1}$ in the Raman spectrum of SLG (Fig. 1). The first two peaks occur at ~4600 cm$^{-1}$ and ~4800 cm$^{-1}$ (see magnified peaks in Fig. 1), and are assigned to combinations of the D and 2D' (D+2D'), and G and 2D' (G+2D') peaks, respectively. These peaks have previously been assigned to the D+2G[15, 16] and 3G[13] peaks, respectively. Tan *et al.* did observe a peak at 4800 cm$^{-1}$ in SWNTs (with 488 nm laser excitation) and assigned it to the G+2D' peak.[14] As mentioned above and also seen in the spectra in Fig. 1, we do not observe the overtone of the G peak (2G), which occurs at ~3160 cm$^{-1}$. In addition, the expected sum of peak frequencies for the D (~1345 cm$^{-1}$) and 2D' (~3232 cm$^{-1}$) peaks is ~4577 cm$^{-1}$, which is ~50 cm$^{-1}$ higher than the sum of frequencies of the D and 2G peaks. Our observed peak frequency in SLG is ~4600 cm$^{-1}$, which is closer to the sum of frequencies for the D and 2D' peaks. Hence for the above reasons, we assign the two peaks at ~4600 and ~4800 cm$^{-1}$ to the D+2D' and G+2D' peaks, respectively. The next weak intensity mode in SLG appears at ~5330 cm$^{-1}$, and is the fourth harmonic of the D peak. Note that the occurrence of a three-phonon (3D) peak is improbable due to the requirement of momentum conservation in the scattering process.[13] Thus the next observable overtone of the D peak phonon is the four-phonon overtone of the D peak (four iTO phonons with equal and opposite momentum) at ~5330 cm$^{-1}$ and is called the 4D peak.[14, 16, 26]

The final four-phonon peak appears at ~5850 cm$^{-1}$ and has been previously assigned to a combination of 2G and 2D phonons in graphite and SWNTs.[14, 16] This peak is very weak in intensity and is shown magnified 50 x in Fig. 1. In order to be consistent with our reasoning concerning the 2G and 2D' peaks above, and because the intensity of the peak at 5300 cm$^{-1}$ is very weak, we assign the four phonon peak at ~5850 cm$^{-1}$ to the 2D'+2D mode. Due to increasing background scattering from the SiO$_2$ substrate, higher order combination modes beyond 6000 cm$^{-1}$ involving 5 or 6 phonons are difficult to observe in our samples. Furthermore, the weak intensity three and four phonon modes above 4500 cm$^{-1}$ are only observed from SLG and very difficult to resolve in multi-layered graphene samples. This is due to larger resonance enhancement for phonon modes in SLG. For example, the high intensity 2D peak in SLG has been attributed to occur due to a triple resonance process, where all the steps in the typical double resonance process are resonant.[2] Thus one would expect the intensities of higher order combination modes involving the 2D phonon to also be higher than the corresponding peaks in BLG, FLG, and HOPG samples. This might explain why the higher order combination modes are observed in SLG while they are very weak in intensity in multi-layer graphene. All the multiphonon peaks described above are listed in Table 1 along with peak assignments and expected frequencies (based on the sum of the individual components). Also included in Table 1 are experimentally observed multiphonon peaks in HOPG for comparison.

In summary, we have observed multiphonon Raman modes (3 and 4 phonon modes) in SLG, BLG and FLG samples on $SiO_2$ substrates. Peak assignments are made based on the dispersion of the peaks versus laser excitation energy, as well as the expected frequency obtained from the sum of the individual components. The D+2D and G+2D peaks frequencies are observed have an almost linear dependence on the number of graphene layers, suggesting an influence of interlayer van der Waals interactions on the peak frequencies. The distinct layer-dependence of the G+2D peak frequency provides another metric for the correct identification of the number of layers in graphene samples. Higher order four-phonon modes are also observed in SLG, while these peaks are much weaker in intensity in graphene samples with 2 or more layers due to the large resonance enhancement in SLG. The peak frequencies of all the combination modes in SLG are also shifted relative to the same modes observed previously in SWNTs; for example the G+2D peak in SWNTs has been observed at ~4300 cm$^{-1}$ (with $E_{laser}$ = 2.41 eV),[13, 14] while the same peak in SLG occurs at ~4260 cm$^{-1}$ with the same laser excitation. This confirms the uniqueness of the phonon band structure of this one-atom thick two-dimensional material, which is different from both the one-dimensional SWNTs and three-dimensional graphite.

ACKNOWLEDGEMENTS

RR is grateful for funding from AFOSR and the National Research Council Associateship program, and thanks Humberto Gutierrez for assistance with part of the Raman spectroscopy measurements. DT, JK, and MI are supported by the National Science Foundation under Grant No. 0955625.

Ref. 3). The few layer graphene samples typically consisted of 3-5 layers. Micro-Raman characterization using 325, 488, and 532 nm laser excitation was performed with a Renishaw inVia Raman microscope.

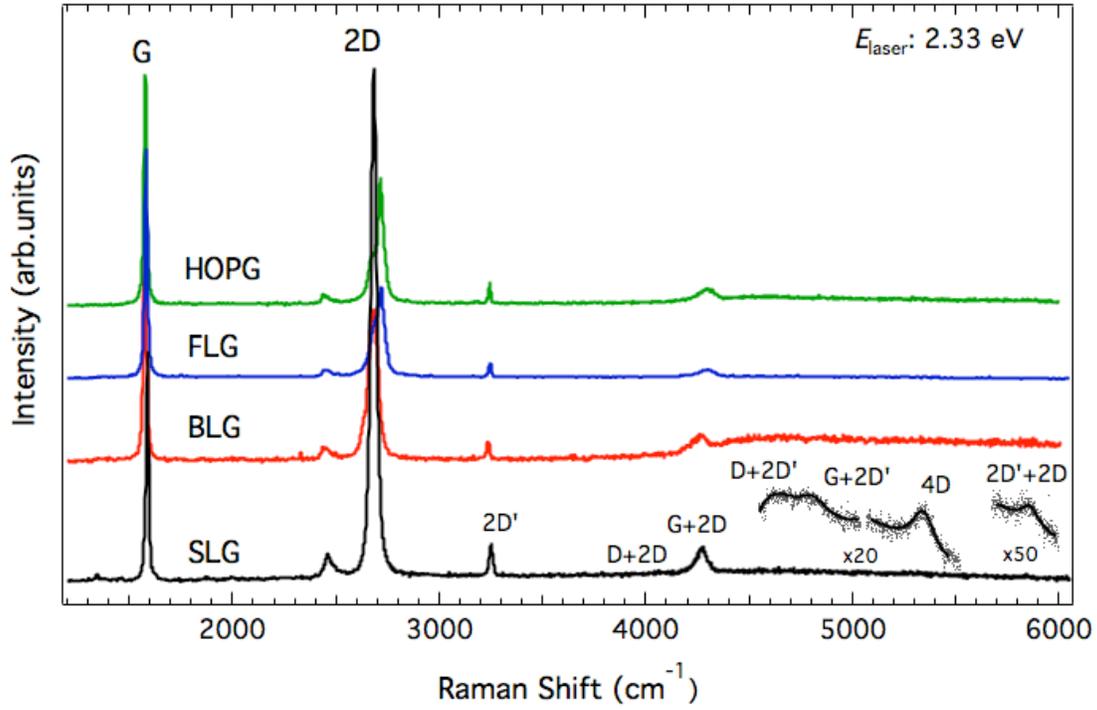

FIG. 1: Higher order Raman spectra from graphene samples collected with $E_{Laser}$ = 2.33 eV. All spectra have been normalized by the G peak intensity. Some of the weak intensity combination modes appearing above 4500 cm$^{-1}$ are magnified for clarity.

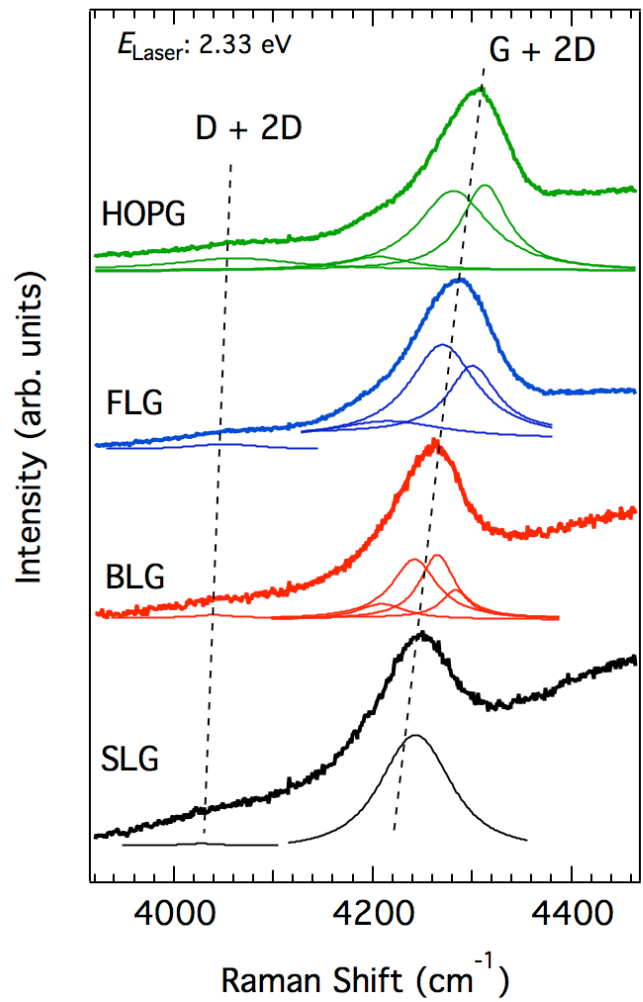

FIG. 2: Higher order combination modes between 4000 – 4400 cm$^{-1}$ in graphene samples collected with $E_{Laser}$ = 2.33 eV. All spectra have been normalized with respect to the G+2D peak intensity for clarity and fitted with Lorentzian peaks. The D+2D and G+2D peaks upshift in frequency with increasing layers.

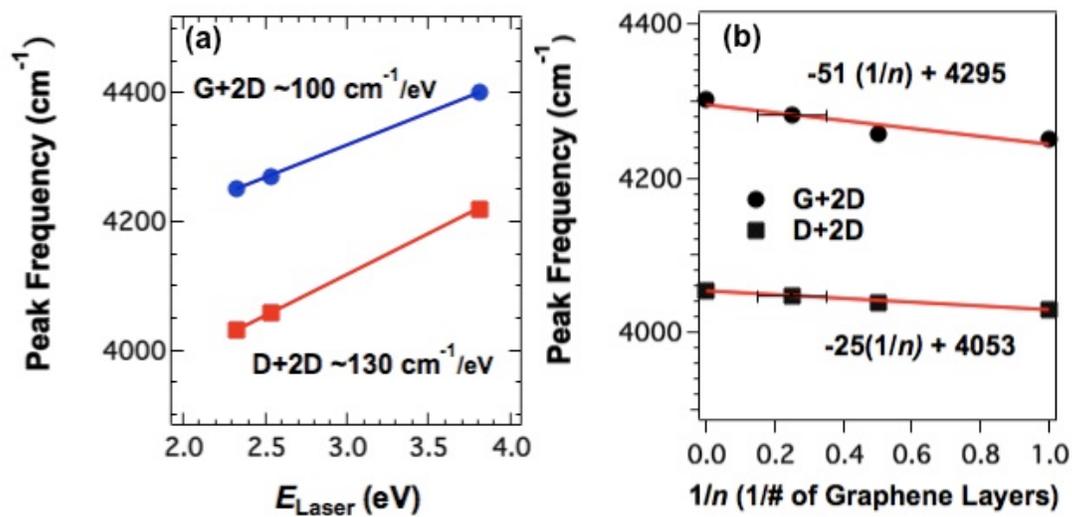

FIG. 3: (a) Dispersion of the G+2D and D+2D peaks in SLG versus laser energy. (b) Peak frequencies of the G+2D and D+2D peaks versus 1/n. The error bars for the FLG samples were obtained from AFM measurements, which confirmed the presence of 3-5 layers.

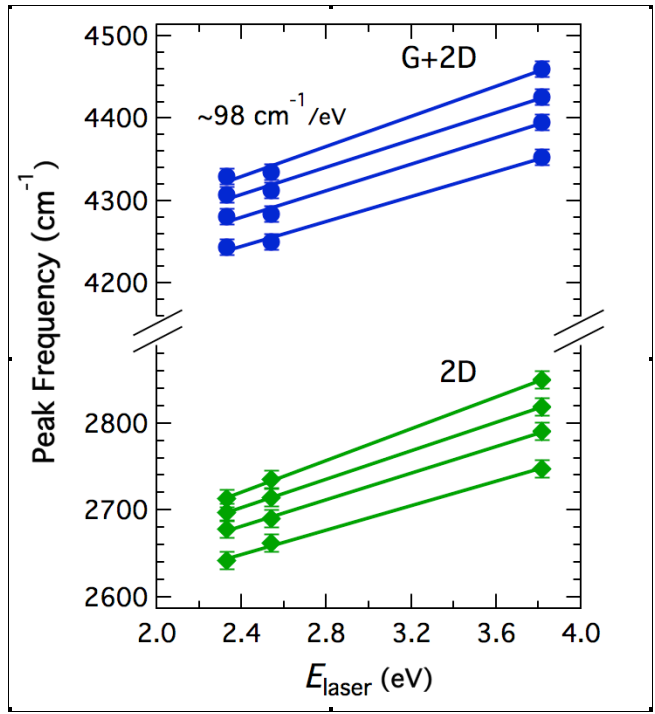

FIG. 4: Dispersion of four components within the G+2D and 2D peaks in BLG versus laser energy. The dispersion of the highest frequency component in the G+2D peak (~98 cm$^{-1}$/eV) is indicated in the figure.

TABLE 1: Peak frequencies and assignments for the multiphonon modes observed in SLG ($E_{Laser}$ = 2.33 eV) and experimentally observed multiphonon peaks in HOPG ($E_{Laser}$ = 2.41 eV) taken from Ref [9].

| Peak Frequency (cm$^{-1}$) | Peak Assignment | Expected Frequency (cm$^{-1}$) | Corresponding peaks observed in HOPG (from Ref. [9]) |
|---|---|---|---|
| 1345 | D | 1345 | 1352 |
| 1583 | G | 1580 | 1580 |
| 1620 | D' | 1620 | 1622 |
| 2675 | 2D | 2690 | 2705 |
| 3232 | 2D' | 3240 | 3244 |
| 4030 | D+2D | 4035 | 4031 |
| 4250 | G+2D | 4270 | 4288 |
| 4600 | D+2D' | 4585 | 4544 |
| 4800 | G+2D' | 4820 | 4830 |
| 5330 | 4D | 5380 | 5370 |
| 5857 | 2D'+2D | 5930 | 5870 |